# DYNAMICAL DERIVATION OF BODE'S LAW

by

Robert W. Bass
Prof. of Physics & Astronomy, BYU, 1971-81
Registered Patent Agent, P.O. Box 1238, Pahrump, NV 89041-1238

## ABSTRACT

In a planetary or satellite system, idealized as *n* small bodies in initially coplanar, concentric orbits around a large central body, obeying Newtonian point-particle mechanics, *resonant* perturbations will cause dynamical evolution of the orbital radii *except* under highly specific mutual relationships, here derived *analytically* apparently for the first time. In particular, the most stable situation is achieved (in this idealized model) only when each planetary orbit is roughly twice as far from the Sun as the preceding one, as observed empirically already by Titius (1766) and Bode (1778) and used in both the discoveries of Uranus (1781) and the Asteroid Belt (1801).

Simplifying the problem by reformulating it as a *hierarchical sequence* of [unrestricted] 3-*body problems*, in which gravitational interactions are ignored except between the central body and the body of interest and the next outwardly orbitally adjacent body, it is proved that the resonant perturbations from the outer body will destabilize the inner body [**& conversely**] *unless* its mean orbital radius is a *unique and specific* multiple $\beta$ of that of the inner body (and, presaging analogous results using elliptical rather than circular generating [decoupled] solutions, wherein the lines of apsides must be either parallel or perpendicular in resonant configurations, there must be a *phase shift* which is an integral multiple of $\pi/2$). In this way a sequence of concentric orbits can each stabilize the next adacent inner orbit, until only the outermost orbit remains; but it is already tied to the collection of inner orbits by conservation of total angular momentum and so the entire configuration becomes stabilized. Let $\mu = M_{MAX}/M$ denote the ratio of the mass $M_{MAX}$ of the largest small body to that, $M$, of the large central body; in our Solar System, $\mu$ is less than $10^{-3}$. Expanding $\beta$ in a power series in $\mu$, the lowest-order terms for **distal multiplier** $\beta$ and **phase shift** $\phi$ are found to start with the following **universal constants** [for $m = 1, 2, 3, \cdots$]:

$$\beta = \beta_O + \cdots \equiv 1/\sqrt{(3/2)^{2/3} - 1} + \cdots = 1.794980 + \cdots, \qquad \phi = m \cdot (\pi/2) + \cdots,$$

which agrees with the observed distal ratios between Jupiter and the Asteroids, and between Saturn & Jupiter, and Uranus & Saturn. In view of Kepler's Law this implies that the **stabilizing resonances** *must be* of the type 2:1 or 5:3 or 3:1 (to the extent that only 3-body interactions and only integers not exceeding 5 are considered). Only analytic complexity has inhibited the present theory from being extended to 4-body and 5-body interactions, as a result of which the observed resonant structure of the Solar System [Ovenden & Roy; Molchanov] can be understood dynamically as the inevitable result of *resonant* perturbational evolution. **Computer simulations** of the Solar System **demonstrating visually** clear lack of stability within a century or two when Bode's Law is significantly violated provide evidence of apparently incontrovertible plausibility of the conjecture here claimed to be proved analytically for the first time as a rigorous **Theorem** pertaining to the new *natural constant* $\beta_o = 1.79498012488927...$, which is *independent* of the Newton-Cavendish parameter $G$ or particular values of the planetary masses.

# Introduction

The present investigation is undertaken as research in dynamical astronomy rather than pure mathematics; accordingly the problem will be presented with the standard of rigor found acceptable in the former, although sufficient precision will be employed that mathematicians interested in perfect rigor should find no difficulty in extending this work both in breadth and depth.

Newton's law of gravitation is only an idealization valid within certain domains. For example, if two point-particles at rest are freed, and only Newtonian gravitation considered, they will accelerate toward each other and collide with infinite speed! On doing the exercise it appears that they will be moving **faster than light** when within their respective Schwarzschild radii ('black hole' radii) of each other. Consequently it is only in the domain of pure mathematics that the *strict* Newtonian *n*-body problem makes even idealized physical sense.

In regularizing the 3-body problem, the mathematician Sundmann was led to postulate that after a collision, the bodies would *rebound* elastically! In such a universe, he proved that if the the initial angular momentum is not zero, then there is a 'regularizing' parameter $\tau$, $-1 < \tau < 1$, such that the Newtonian time $t = t(\tau)$ and positions $x^i = x^i(\tau)$ and velocities $v^i = v^i(\tau)$ are all **analytic** functions of the **complex** variable $\tau$ on the unit disk $|\tau| < 1$, taking real values on the real axis, and with $t$ going from $-\infty$ to $+\infty$ as $\tau$ moves from -1 to 1. While a justly celebrated triumph of mathematics [1], [2] this result has no physical meaning, and it has been proved that Sundmann's series converge so slowly that for practical use a computer bigger than the Solar System would be required!

Therefore, recognizing that mathematical models are only *models* and not reality, it seems legitimate to employ physical intuition to simplify the model to the point that it can be solved in the researcher's lifetime.

This author is fully aware of W.M. Smart's important result that *if* the traditional perturbation series of dynamical astronomers do happen to converge (which, thanks to Poincaré [3], we know that *in general* they do **not**), and *if* the third term's size is any indication of the rate of convergence of the first two terms, then such series cannot be trusted for more than about '200 years' [4]. Accordingly, he is completely unconvinced by the various attempts to integrate the solar system forwards and backwards in time numerically for hundreds of millions of years; indeed, Roy, one of the leading participants in such projects, has given a ten-item checklist [5, p. 266] of physical effects that could have altered the results but which were neglected, and an additional 13-item checklist [*id.*, 267] of numerical problems which have not been laid to rest. Accordingly the intellectual honesty of such recent reports as [6] is admirable: it is specifically admitted that the results probably have only 'qualitative' rather than quantititative significance -- for example, the **NEA** (Near Earth Asteroid) **1991 TB**$_2$, when integrated forward in time subject to the resonant perturbations of Jupiter, had its eccentricity 'pumped' all the way from zero to unity in a mere 30,000 years, after which it fell into the Sun -- but that same asteroid, when integrated backwards in time, was found to to *emerge* [sic!] from the Sun and be ejected into its present position between Mars and Jupiter some 2,**5**00,000 years in the past! (Table 1, p. 317, [6].)

Similarly, a "numerical simulation ... of a real object [asteroid **1991 VP5**] in the 5:2 commensurability, when the full solar system is taken into account" shows [30] that in as little as 12,500 years the eccentricity can be pumped from about 0.1 to more than 0.4 (sufficient to cross the orbit of Mars) or even 0.625 (sufficient to become an Earth-crosser)!

Finally, just-published simulations of Jupiter-sized planets orbiting hypothetical stars [33], which demonstrate that eccentricities can be pumped up or down resonantly to permit near collisions (or ejection) within centuries rather than eons, have moved MIT astrophysicist Philip Morrison [31] to comment in the latest *Scientific American* that "few orbital radii within our sun's system remain vacant where additional planets could permanently circle... Mercury may...become next...to depart!"

These three 1994-1996-published specific examples illustrate the tremendous potential importance of *resonance* for understanding of the 3-body problem (hence, *a fortiori*, for the general *n*-body problem).

However, it has been a highly controversial question as to whether or not the observed resonances in our solar system, and in the known natural satellite systems, are meaningful or coincidental. Goldreich [7], [8], [9] seems to have definitively settled the matter in the case of the satellites, having pointed out that in the case of nearby bodies tidal forces can cause dynamical evolution to occur, but that *certain particular* resonances are stabilized by tidal friction, and so will be evolved into and thereafter remain, apparently having been in the corresponding idealized Newtonian [conservative, Hamiltonian] orbits forever. One consequence of the present paper is proof that Goldreich's idea is applicable even in the absence of tidal friction (although such non-conservative forces can "lock" a resonance which has been reached via strictly Hamiltonian forces).

Meanwhile, the subject of dynamical *chaos* (originated by Poincaré [3] and well known in the 1920's to his Chicago successor Moulton [10] -- who specifically proved that the period of a spherical pendulum is a *discontinuous* [!] function of the pendulum's initial conditions) has become more widely appreciated. There are now many physicists who understand why the Golden Mean ($\sqrt{5}$ -1)/2 = 0.6180 ⋯ is quite literally the most *irrational* of all possible irrational numbers [11], [12], and even layman have heard on TV that weather dynamics has such *extreme sensitivity* to initial conditions that 'a butterfly flapping its wings in South American can cause a tornado in North America' [13]. One therefore expects dynamical astronomers to know that between any two numbers (rational or irrational) another of the opposite kind may be found, i.e. both species of numbers are *dense* on the real line (-∞, +∞). Those who have forgotten that in the sense of probability theory throwing darts at the real number segment [0,1] = { $r$ | 0 ≤ $r$ ≤ 1 } yields the probability *zero*

that *r* be rational and the probability *unity* that *r* be irrational can brush up in **Appendix 1** below, which also disposes of the canard that resonant orbits, having probability zero, may safely be ignored.

What then can be the *physical* significance of a *resonance* or *commensurability* between planetary angular velocities $\omega_i$, $\omega_j$, defined to be the existence of [usually] small, **mutually prime** positive integers $m_i$, $m_j$ such that

$$m_j \cdot \omega_i - m_i \cdot \omega_j = 0, \tag{1}$$

other than to guarantee the existence of an *underlying* frequency $\omega$ such that

$$\omega_i/m_i = \omega_j/m_j := \omega, \qquad \omega_i = m_i \cdot \omega, \qquad \omega_j = m_j \cdot \omega, \tag{2}$$

in terms of which the two motions now have a commmon *period* $T := 2 \cdot \pi/\omega$, wherein one body executes exactly $m_i$ revolutions while the other body executes exactly $m_j$ revolutions, after which the initial configuration is attained, and then the motion repeats itself endlessly?

Anyone who has pushed a child on a swing knows that an *external* force, no matter how small, if in **synchonicity** with a *natural period* of an autonomous dynamical system, can eventually 'pump' the amplitude of oscillation to arbitrarily large magnitudes.

This is perfectly illustrated in the cited cases of **NEA 199**1 **TB**$_2$ and **1991 VP5**.

Dynamical astronomers since Laplace have cited the 'long period' or 900-year oscillation in the orbit of Saturn caused by its 5:2 resonance with Jupiter. However, some 2:1 orbital resonances (e.g. at perihelion [14]) are known to be destabilizing, while others (e.g. at aphelion) are known to be stabilizing.

Moreover, the question of dynamical stability of a given periodic motion is an order of magnitude more difficult mathematically [15] than the mere question of whether or not such periodic motion exists to begin with. Nevertheless, an initial effort will be made in what follows to demonstrate that both subjects can be studied simultaneously with dramatically enlightening results.

Most current basic research in analytical celestial mechanics centers on the celebrated *KAM* theorem [16], [17], [18], which has attained success only by *avoiding* the pesky subject of resonances. When this author first heard of these results in the mid 1960's, his initial reaction was that they represent the largest mathematical advance in celestial mechanics of the past century. Suppose now that the two angular frequencies in question are *incommmensurable*, in the sense that their ratio is an **irrational** real number *r*:

$$\omega_i/\omega_j = r. \tag{3}$$

Based on some ideas of Siegel [19], ultimately stemming from Poincaré (who considered the case that $r^2 = p/q$ for relatively prime small integers *p*, *q*), Kolmogorov outlined a method of 'proving the stability of the solar system', which was completed by his student Arnol'd and, independently, by Siegel's former student Moser [18], [2]. One says that a number *r* is *strongly irrational* if there is another number ε > 0 such that

$$| r - (p/q) | \geq \varepsilon/|q|^2 \tag{4}$$

for all possible pairs of relatively prime integers (*p*,*q*) whatsoever. It can be proved that such numbers are also **dense** on the real line (-∞, +∞), and that on any fixed, finite interval, in the sense of **advanced probability theory** ['measure' theory], such numbers have a **'positive probability'** [positive measure], which tends to unity as ε tends to zero [*cf.* Appendix 1 below].

This concept of the **strong incommensurability** of two frequencies can be generalized to that of strong incommensurability of *N* frequencies:

$$| m_1 \cdot \omega_1 + \cdots + m_N \cdot \omega_N | \geq \varepsilon/(| m_1 | + \cdots + | m_N |)^{N+1}, \tag{5}$$

with similar results regarding the **density** and **positive probability** in real *n*-dimensional Euclidean frequency-space $\mathbb{E}^n$. In the general Newtonian *n*-body problem it was first showed in 1876 by Simon Newcomb, in probably the deepest piece of mathematics hitherto done in the Western Hemisphere, that one should take $N = 3 \cdot n - 1$ (or, for the planar problem, $N = 2 \cdot n$); what Newcomb proved [20] is that there is a *formal* multiply-periodic Fourier Series, not necessarily convergent, which satisfies the Newtonian dynamical equations when compared term-by-term. Poincaré, who subtitled Vol. 2 of the original French edition of his book [3] with the name of Newcomb, was moved to discover that by making infinitesimal changes in the Fourier coefficients of each term in the Newcomb series, similarly to the covering of all rationals exhibited in Appendix 1 below, he could convert the series into one which assuredly diverged; therefore, it is impossible in general for Newcomb's series to converge *uniformly*; when they do, one has **orbital stability**. What the *KAM* trio achieved was to prove that under certain circumstances, when the frequencies are strongly incommensurable, Newcomb's series *do* converge; but the convergence is not uniform, and consequently there may be, arbitrarily close to the initial conditions leading to a *KAM* quasi-periodic motion, an ejection-collision **'wild'** motion.

In contrast, the series associated with resonant motions *can* converge uniformly, as do those proved to exist in the present work; therefore, the presently-constructed resonant motions are *a priori* orbitally stable. (It is notorious that Newcomb's series in the case of a resonance, leading to the famous 'problem of small divisors' [1]-[2], may also, in the circumstances excluded from the present theory, diverge exponentially because of numerators which contain higher and higher powers of expressions like the left-hand side of (5), which are either *arbitrarily* close to zero [contradicting (5)], or exactly zero, leading to instant divergence unless the corresponding Fourier coefficient is somehow guaranteed *a priori* to vanish simultaneously; such simultaneous vanishing is the consequence of the novel analytical conditions to be derived herein.

Applied to a Solar System, what the *KAM* theory proves is this: consider a planetary configuration of *n* small masses in orbit around a much larger central mass. Suppose that the initial inclinations and eccentricities are 'small' (i.e. when 'sufficiently small' the theorem applies, but no quantitative estimate is given for how small these small quantities must be), and that the initial 3·*n* frequencies of the 3 angular variables in each of the *n* bodies' 6 astronomical elements are strongly incommensurable. Then there is a positive probability that arbitrarily close to that configuration there is an actual solution of the Newtonian ($n + 1$)-body problem which is *multiply periodic* (often called *quasiperiodic*), i.e. which endures without close approaches or near collisions from eternity to eternity.

Unfortunately, the detailed investigations of the *KAM* school demonstrated that also, arbitrarily close to the phase-space multi-tori on which the quasi-periodic motions are found, there are what Poincaré called *wild motions* which include the infamous *heteroclinic* points that are so mindbogglingly complex that it takes a chapter of an advanced book to begin to describe them, and for which one may consult the literature already cited.

Furthermore, even more frustrating to the present author, the wonderfully deep *KAM* theory is silent regarding configurations of true resonance, and therefore appears inapplicable, even in principle, to observed dynamical resonances among the planets and satellite systems.

In 1974 this author summarized these matters in two papers in *Pensée* [21], [22] (and one in *Kronos* [23]). Archie Roy quoted extensively (with appreciated explicit citation & acknowledgement) from these papers in the final two pages of his Chapter on the **Stability of the Solar System** in his authoritative monograph on *Orbital Motion* [5]. The way I left it in 1975 in what Roy calls my 'concise summary' is that a resonant motion can be either orbitally stable or unstable and that no general rule is known, i.e. each case must be investigated individually. (In a 1991 *New Scientist* article, Carl Murray stated without citation or proof other than an intuitive sequence of drawings that when $n = 2$, stability or instability in a resonant configuration depends upon whether or not the periodically repeated conjunctions are at perihelion or aphelion of the inner planet [14].)

I am now extraordinarily indebted to David Talbott (publisher of *Pensée* and its forthcoming successor *Search*) for stimulating me to return, after 20 years, to a reconsideration of the **physically all-important** case left unresolved by the *KAM* theory.

The basic strategy in the following is as follows: the famous "Poincaré map" which is now often implemented by physicists using numerical integration [34] will be implemented analytically. In the present context, this means that an arbitrary line-segment will be defined transverse to a periodic generating orbit, and nearby orbits will be followed until this 1-dimensional "surface of section" is intersected a second time. This defines a continuous mappng of the section into itself, of which map the initial periodic solution is now a fixed point. Behavior of nearby orbits can now be studied by considering the simpler problem of iterating this map in a small neighborhood of the fixed point. It will be proved below that no fixed point exists (i.e. no periodic solution exists) unless at $\mu = 0$ the distal ratio defined in the Abstract is precisely $\beta_o$.

## Problem Formulation

Let $x^i \in \mathbb{E}^2$ ($i = 1, 2, 3, \cdots, n$) denote 2-vectors or elements of 2-dimensional real Euclidean space $\mathbb{E}^2$ which represent the positions of $n$ point-particles of masses $M_i$ in orbit around a central body of mass $M_o$ and position $x^o$. Let $v^i \in \mathbb{E}^2$ denote their velocities $\dot{x}^i \equiv dx^i/dt$ with respect to time $t$, where $\dot{} := d/dt$. Let $x \cdot y \equiv (x,y)$ denote the *scalar product* between any two vectors $x$, $y$, and let $\|x\| := (x,x)^{1/2}$ denote the *norm* (length) of any vector.

It is assumed that $\mu_i := M_i/M_o \ll 1$. Now rescale the masses so that $M_o = 1$ and the smaller masses are $\mu_i \ll 1$. Let $\mu_{MAX}$ denote the largest of the small masses, and rewrite them as

$$\mu_i = \mu \cdot \mu_i^o = \mu \cdot \varepsilon_i^o \cdot \mu_{MAX}, \quad 0 \leq \mu \leq 1, \quad (0 < \varepsilon_i^o \leq 1), \tag{6}$$

so that $\mu$ is the *perturbation parameter*. All that will be PROVED here is for $\mu$ 'sufficiently small', though there are good reasons for believing that an analytical continuation (in the manner of Poincaré) can be made all the way from $\mu = 0$ to $\mu = 1$; this is because the **Leray-Schauder Index** of the 'generating solution' at $\mu = 0$ will be *proved* to be unity, and the fact that this *integer-valued* topological invariant, specifying in some sense the true multiplicity of actually existing solutions, is a *continuous* homotopy invariant and so can change its value *discontinuously*, i.e. terminate', ONLY at a bifurcation or singularity of the solution (i.e. a collision or an ejection), demonstrates that the homotopy on $\mu \in [0,1]$ is legitimate *unless* there is an intermediate value of $\mu < 1$ at which there is a collision between two bodies or one body is ejected to infinity; this makes rigorous Strömgren's *Principle of Natural Termination*, discovered empirically by numerical integration [1], in which a family of periodic solution*s* being studied by the variation of a parameter can cease to exist *only* at an ejection/collision event.

Now introduce *relative* coordinates, in which each $x^i$ is replaced by $(x^i - x^o)$, although for convenience the latter will be renamed as $x^i$; it is well known that then the system becomes, for the indices $i = 1, 2, \cdots, n$,

$$\dot{x}^i = v^i, \tag{7a}$$

$$\dot{v}^i = -G \cdot (1 + \mu_i) \cdot (x^i/\|x^i\|^3) - G \cdot \sum_{\substack{j=1 \\ j \neq i}}^{n} \mu_j \cdot \left\{ [x^j/\|x^j\|^3] - [(x^j - x^i)/\|(x^j - x^i)\|^3] \right\}, \tag{7b}$$

where the absolute acceleration of the coordinate system's origin, which now coincides with the position of the Sun or central body, is the source of the first term in the summation (cf. Kurth [24], pp. 84-85), and where $G$ denotes the Newton-Cavendish parameter or gravitational constant.

Now, following Kurth, assume that each planet is affected significantly only by the Sun and the next further outward planet:

$$\dot{v}^k = -G \cdot (1 + \mu_k) \cdot (x^k / \|x^k\|^3) + G \cdot \mu_k \cdot \left[ [(x^m - x^k)/\|x^m - x^k\|^3] - [x^m/\|x^m\|^3] \right],$$

$$k = (i,j), \quad j = (i+1); \quad m = m(k); \quad m(i) = j; \quad m(j) = i; \quad (i = 1, 2, \cdots, n-1). \tag{7c}$$

Again, following Kurth, I shall solve the Newtonian system (7a,c) by successive approximations, but in a modified manner. Note that if $\mu_k = 0$, then (7c) can be solved by concentric circular solutions, with frequencies $\omega_k$ given by **Kepler's Third Law** (1619) as

$$(\omega_k)^2 = G \cdot (1 + \mu_k)/(\rho_k)^3, \qquad \rho_k := \|x^k\|. \tag{8}$$

For the purpose of successive approximations, replace (7c) by the equivalent *ordinary differential equation* (**ODE**)

$$\dot{v}^k = -(\omega_k)^2 x^k + G \cdot \mu_k \cdot \left[ [(x^m - x^k)/\|x^m - x^k\|^3] - [x^m/\|x^m\|^3] \right] + [(\omega_k)^2 - G \cdot (1 + \mu_k)/\|x^k\|^3] \cdot x^k, \tag{9}$$

which makes clear the fact that the system is just a perturbation of the *harmonic oscillator* problem.

We shall need the following lemma, which is an obvious modification of well known basic results in the theory of ODEs [25], [26], [27], [28].

## Lemmas from ODE Theory

<u>DEFINITION</u>. *An m-vector function f: $\mathbb{E}^m \to \mathbb{E}^m$ has a global Lipschitz constant $\kappa$ if there is a positive real number $\kappa > 0$ such that, for all $z^i \in \mathbb{E}^m$,*

$$\| f(z^2) - f(z^1) \| \leq \kappa \cdot \| z^2 - z^1 \|. \tag{10}$$

<u>THEOREM</u>. *Consider the m-vector ODE Initial Condition Problem* (**ICP**)

$$\dot{z} = f(z), \qquad z(0) = z^o, \qquad (0 \leq t \leq T), \tag{11}$$

*where f satisfies a global Lipschitz condition, and where the maximum time T is arbitrary but finite. Let A denote an arbitrary $m \times m$ constant real matrix, and define the vector function g by $g(z) := f(z) - Az$, which also is globally Lipschitzian, with Lipschitz constant now $[\kappa + \|A\|]$,*

*where* $\|A\|$ *denotes the Euclidean norm of A. Now construct the sequence of functions* $\{z^k(t)\}$ *by successive solution of the forced linear ODE*

$$\dot{z}^k = Az^k + g(z^{k-1}(t)), \quad z(0) = z^o, \quad (k = 2, 3, \cdots), \tag{12}$$

*where the initial iterate* $z^1(t)$ *is any arbitrary continuous function of t on* [0, T] *such that its initial value* $z^1(0) = z^o$*; or, equivalently, construct the sequence by*

$$z^k(t) = exp(At) \cdot z^o + \int_0^t exp(A[t - \tau]) \cdot g(z^{k-1}(\tau)) \, d\tau, \quad 0 \le t \le T, \tag{13}$$

*i.e. by integrating the right-hand-side to obtain the left-hand side, and then inserting the result into the right-hand-side and repeating the operation. The resulting sequence is guaranteed to converge uniformly on* [0, T] *to the exact and unique solution of* (11)*, no matter what is the arbitrary initial iterate* $z^1(t)$*, and no matter what is the arbitrary matrix A, and no matter how large is the [fixed] time-interval* [0,T]*.*

PROOF. Any standard work on ODE's will show how to use Lagrange's method of *variation of constants* to prove that the ODE ICP

$$\dot{z} = Az + g(z), \quad z(0) = z^o, \quad (0 \le t \le T), \tag{14}$$

is completely equivalent to the Volterra integral equation problem

$$z(t) = exp(At) \cdot z^o + \int_0^t exp(A[t - \tau]) \cdot g(z(\tau)) \, d\tau, \quad 0 \le t \le T, \tag{15}$$

where

$$exp(At) := \sum_{k=0}^{\infty} A^k \cdot t^k / k!. \tag{16}$$

Therefore, if the sequence in (12)-(13) converges uniformly, then the result satisfies (15) and so is the solution of (11). Because every continuous function on a closed and bounded subset of a finite-dimensional Euclidean space has (and assumes) a finite maximum and minimum, there exists (on the closed, bounded subset $[0,T] \subset \mathbb{E}^1$)

$$\phi_o = \underset{t \in [0,T]}{MAX}\{\|z^1(t) - z^o(t)\|\}. \tag{17}$$

By repeatedly applying the triangle inequality to (16) one finds readily that

$$\|exp(At)\| \le \sum_{k=0}^{\infty} \|A\|^k \cdot t^k / k! \equiv exp(\|A\| \cdot t) \le \gamma := exp(\|A\| \cdot T). \tag{18}$$

Accordingly it is easy to prove by induction that (setting $\kappa_1 = \kappa + \|A\|$)

$$\| z^{k+1}(t) - z^k(t) \| \leq \phi_o \cdot (\gamma \cdot \kappa_1 \cdot t)^k / k!, \tag{19}$$

$$\lim_{N \to +\infty} \left\{ \| z^N(t) - z^o(t) \| \right\} \leq \sum_{k=0}^{\infty} \| z^{k+1}(t) - z^k(t) \| \leq \phi_o \cdot exp(\gamma \cdot \kappa_1 \cdot T), \tag{20}$$

for each $t \in [0,T]$, so that $\{ z^k(t) \}$ is a Cauchy sequence (for each $t$) and its limit $z(t)$ must exist. Consideration of the approximating sums $\Sigma_N$ and $\Sigma_{N+1}$ to the series in (20) shows that the convergence is uniform. □

In the sequel, we shall frequently use the 2 × 2 identity matrix $I_2$ and its 'imaginary' skew-symmetric counterpart $J_2 \equiv -(J_2)'$ defined as

$$I_2 = \begin{bmatrix} 1 & 0 \\ 0 & 1 \end{bmatrix}, \quad J_2 = \begin{bmatrix} 0 & 1 \\ -1 & 0 \end{bmatrix}, \quad (J_2)^2 = -I_2. \tag{21}$$

For present purposes, the result that

$$exp(J_2 \cdot \phi) \equiv \begin{bmatrix} cos(\phi) & sin(\phi) \\ -sin(\phi) & cos(\phi) \end{bmatrix}, \tag{22}$$

will be all-important. (To prove it, insert (21c) into (16).) A similar result which will be needed is that if the 4 × 4 matrix $A$ is defined by (using *MATLAB* notation) $A = [0, I_2; -\omega^2 I_2, 0]$, then it is easy to prove by induction that

$$A^{2k} = (-1)^k \cdot \omega^{2k} \cdot I_4, \quad A^{2k+1} = (-1)^k \cdot \omega^{2k} \cdot A, \tag{23}$$

whence from (16) it is immediate that [letting $I$ denote $I_2$]

$$exp(A \cdot \phi) = exp\begin{bmatrix} 0 & I \cdot \phi \\ -\omega^2 \cdot I \cdot \phi & 0 \end{bmatrix} = \begin{bmatrix} cos(\omega \cdot \phi) \cdot I & [1/\omega] \cdot sin(\omega \cdot \phi) \cdot I \\ -\omega \cdot sin(\omega \cdot \phi) \cdot I & cos(\omega \cdot \phi) \cdot I \end{bmatrix}. \tag{24}$$

## Principal Result

THEOREM. *Consider the semi-restricted Copernican-Newtonian* $(n + 1)$-*body problem as formulated in* (7a) *and* (9), *with Initial Conditions* (ICs) *corresponding to Keplerian concentric circular orbits, i.e. in which*

$$x^k(0) \cdot v^k(0) = 0, \tag{25a}$$

*so that 3 parameters suffice to specify the ICs. Let these parameters be defined by the triplet* $(\rho, \theta, \omega)$ *given by polar coordinates in which* $\rho$ *denotes the initial orbital radius,* $\theta$ *the initial phase angle, and* $\omega$ *the initial angular velocity, assumed to conform to Kepler's Law* (8), *which* *reduces the IC to a 2-parameter set* $(\rho,\theta)$. *Specifically, assume* (8) *and let the ICs of* (7a) & (9) *be*

$$x^k(0) = \rho_k(cos(\theta_k^o), sin(\theta_k^o))', \quad v^k(0) = \omega_k \cdot J_2 \cdot x^k(0) \equiv -\omega_k \cdot \rho_k(cos(\theta_k^o), sin(\theta_k^o))', \tag{25b}$$

*where* ' *denotes vector-matrix transposition. Suppose further that the ICs are so chosen that the frequencies are resonant as in* (1)*; for later convenience, we may assume that*

$$1 \leq m_i, m_j \leq 5, \tag{26}$$

*although the main result holds when these integers are arbitrarily large, which means that the initial frequencies* $(\omega_i, \omega_j)$ *are essentially arbitrary. Now, keeping the* **resonant** *frequencies* $(\omega_i, \omega_j)$ *found at* $\mu = 0$ **fixed***, consider the variation of* $(\rho,\theta)$ *as a function of* $\mu$ *in such a way as to preserve the periodicity* [namely, "isoperiodic" continuation]:

$$x^i(t + T) \equiv x^i(t), \quad T = 2\cdot\pi/\omega, \quad \omega = \omega_i/m_i = \omega_j/m_j, \tag{27}$$

*while ignoring the question of periodicity of* $x^j$, $j = i + 1$. *A necessary and sufficient condition that for sufficiently small* $\mu$ *there exist* ( $\rho_i(\mu)$, $\theta_i(\mu)$ ) *preserving the periodicity* (27) *is that there exist a phase-shift* $\phi = \phi(\mu)$ *and a distal multiplier* $\beta = \beta(\mu)$ *such that*

$$\theta_{i+1}(\mu) = \theta_i(\mu) + \phi, \tag{28}$$

$$\rho_{i+1}(\mu) = \beta \cdot \rho_i(\mu), \tag{29}$$

*where, in the second post-Keplerian approximation* $(\phi,\beta)$ *are given by*

$$\phi = m\cdot(\pi/2) + \cdots, \quad (m = 1, 2, 3, \cdots), \tag{30a}$$

$$\beta = \beta_o + \cdots \equiv 1/\sqrt{(3/2)^{2/3} - 1} + \cdots = 1.794980 + \cdots, \tag{30b}$$

*whence, by Kepler's Law in the form*

$$r \equiv m_i/m_{i+1} = \beta^{3/2} + \cdots = 2.40 + \cdots, \tag{30c}$$

*the ONLY* [!] *possibilities for the resonance in the low-order case* (26) *are*

$$m_i : m_{i+1} = 2:1 \quad or \quad 5:2 \quad or \quad 3:1. \tag{30d}$$

REMARK. Cancel the arbitrary **supplementary** assumption (26) which leads to (30d); then the initial generating orbits can be placed arbitrarily closely to **ANY** pair of concentric circles with arbitrary radii (because an arbitrary irrational $\beta$ in (29), and so its $(3/2)^{th}$ power $r$ as in (30c), can be approximated as closely as desired by a rational number $m_i/m_{i+1}$). In this way it can be seen that the *unique* distal multiplier β given in (30b) as 1.80 at μ = 0 is a **new natural constant** which is completely independent of the assumed trial value of β at μ = 0 ! Another way to state this is that with **probability unity** the result (30a,b,c) applies to **ANY** planetary 3-body problem of the type under consideration; see Appendix 1 for details of the type of considerations implying that the initial configurations to which the result (30) does **NOT** apply have **"zero probability"** of occurrence!

PROOF. Let $e^i$ denote the columns of $I_2 = (e^1, e^2)$. Then at μ = 0 the solution of the given ODE ICP is

$$x^k(t) = \rho_k \cdot exp(J_2 \cdot \theta_k) \cdot e^1, \quad \theta_k = \omega_k \cdot t + \theta_k^o = m_k \cdot \sigma + \theta_k^o, \quad \sigma := \omega \cdot t. \tag{31}$$

We want this to be the linear part of the reformulation presented in the Lemma, so choose $A$ as in (24), and then re-express the problem in the equivalent form (15), where the required matrix exponential is given by (24); here now $z = (x', v')' \in \mathbb{E}^4$ and, correspondingly, the $g(z)$ in (15) is given by $g = (0', [g^{km}]')'$, where

$$g^{km} = G \cdot \mu_k \cdot \left[ [(x^m - x^k)/\|x^m - x^k\|^3] - [x^m/\|x^m\|^3] \right] + [(\omega_k)^2 - G \cdot (1 + \mu_k)/\|x^k\|^3] \cdot x^k. \tag{32}$$

Thus, define the 'osculating' **generating solution** at μ = 0 by

$$x^{k,o}(t) = cos(\omega_k \cdot t) \cdot x^k(0) + [1/\omega_k] \cdot sin(\omega_k \cdot t) \cdot v^k(0), \tag{33a}$$

$$v^{k,o}(t) = cos(\omega_k \cdot t) \cdot v^k(0) - \omega_k \cdot sin(\omega_k \cdot t) \cdot x^k(0). \tag{33b}$$

Now the problem is rigorously equivalent to solving the integral equation

$$x^k(t) = x^{k,o}(t) + \int_0^t [1/\omega_k] \cdot sin(\omega_k [t - \tau]) \cdot g^{km}(x^k(\tau), x^m(\tau)) \, d\tau; \tag{34a}$$

although the formulation (24) gives a second equation

$$v^k(t) = v^{k,o}(t) + \int_0^t cos(\omega_k [t - \tau]) \cdot g^{km}(x^k(\tau), x^m(\tau)) \, d\tau, \tag{34b}$$

the second is a consequence of differentiating the first with respect to time $t$, and therefore is of no further consequence, after noting that in (33a) the initial velocity $v^k(0) = \omega_k \cdot J_2 \cdot x^k(0)$ is already defined in terms of $x^k(0)$ by hypothesis.

It is a well-known consequence of basic ODE theory that the (necessarily unique) solutions of (7a) and (9) for μ = 0, namely (33), and the solutions for sufficiently small μ > 0 remain arbitrarily close together for any finite time, specifically here for $0 \leq t \leq T = 2 \cdot \pi/\omega$, provided only that μ be sufficiently small. Therefore for sufficiently small μ one knows *a priori* that the solutions remain within planar concentric annuli surrounding the circular orbits (33), and therefore do not approach each other during the time of interest. Accordingly one may find a global Lipschitz constant for such relevant domains of the Cartesian products of $\mathbb{E}^4$ (in which the Jacobian matrix of the right-hand side of (11) is continuous, and so has a bounded norm $\kappa$ in the relevant $z$-domain). Therefore the Lemma is applicable, for sufficiently small $\mu$, and so the solution of (34) exists and can be constructed by successive approximations (with $j = 1, 2, 3, \cdots$) :

$$x^{k,j}(t) = x^{k,o}(t) + \int_0^t [1/\omega_k] \cdot \sin(\omega_k[t - \tau]) \cdot g^{km}(x^{k,j-1}(\tau), x^{m,j-1}(\tau)) \, d\tau, \tag{35}$$

taking as the initial iterate the decoupled Keplerian solution (33), for $0 \leq t \leq T$.

Next, look at the solution of (34) only at every revolution of duration $T = 2\cdot\pi/\omega$, wherein the *commensurability* of $\omega_i$ and $\omega_{i+1}$ ensures that there is a **common period** to the initial iterates of the two adjacent orbits. This give the famous **Poincaré map**:

$$x^k(T) = x^k(0) - \mu \cdot (G\mu_m^o/[m_k \omega^2]) \cdot \int_0^{2\pi} \sin(m_k \sigma) \cdot h^{km}(x^k(\sigma), x^m(\sigma)) \, d\sigma, \tag{36a}$$

$$h^{km} := \left[ [(x^m - x^k)/\|x^m - x^k\|^3] - [x^m/\|x^m\|^3] \right] + [(\omega_k)^2 - Gp \cdot (1 + \mu_k)/\|x^k\|^3] \cdot x^k. \tag{36b}$$

We shall prove that this map can be expressed in the form

$$x^k(T) = x^k(0) - \mu \cdot (G\mu_m^o/[m_k \omega^2]) \cdot f^{km}(\rho_k, \theta_k^o, \rho_m, \theta_m^o; \mu), \tag{37a}$$

$$f^{km}(\rho_k, \theta_k^o, \rho_m, \theta_m^o; \mu) := \int_0^{2\pi} \sin(m_k \sigma) \cdot h^{km}(x^k(\sigma), x^m(\sigma)) \, d\sigma, \tag{37b}$$

where for certain specific values of $(\rho_k, \theta_k^o)$ the Jacobian determinant of $f^{km}$ with respect to $(\rho_k, \theta_k^o)$ is positive at $\mu = 0$, and where $f^{km} = 0$ for these same specific values, at $\mu = 0$; then, by the Implicit Function Theorem there exist functions $(\rho_k(\mu), \theta_k(\mu))$ for sufficiently small values of $\mu$ which satisfy $f^{km} \equiv 0$ and so which provide the desired periodicity of $x^k(T)$.

It will become evident that the conditions to be derived are also necessary, because if they are not satisfied, $f^{km}$ provides a *resonant* forcing term which drives $x^k(T)$ ever farther from $x^k(0)$, and will do so without bound (or until the resonant 'pumping' drives the map's next iterate out of the map's domain of definition).

To anticipate the results of a rather lengthy and arduous calculation, it will be proved below that $f^{km}$ has the form

$$f^{km} = \eta_{km,m} \cdot \exp(J_2 \Phi_{km,m}) \cdot x^m(0) - \eta_{km,k} \cdot \exp(J_2 \Phi_{km,k}) \cdot x^k(0), \tag{38}$$

where the scalar functions $\eta_{kmj}(\mu)$ and $\Phi_{kmj}(\mu)$ have the properties that

- 13 -

$$\eta_{km,m}(0) = 0, \quad \eta_{km,k}(0) = 1/2, \quad \Phi_{km,k}(0) = \pi/2, \tag{39}$$

so that at $\mu = 0$ the Jacobian matrix $H$ of $f^{km}$ with respect to $x^k(0)$ is given by

$$H = -(1/2)\cdot exp(J_2\cdot[\pi/2]) = -(1/2)\cdot J_2, \quad det(H) = 1/4 > 0, \tag{40}$$

and, as claimed, the **Leray-Schauder Index** [29] of the chosen generating solution is unity.

It is evident by inspection that upon the second iteration the third term (36b) drops out by the choice of a Keplerian generating solution, so that for consideration of the second iterate of the successive approximations we may simplify (36b) to

$$h^{km} := [(x^m - x^k)/\|x^m - x^k\|^3] - [x^m/\|x^m\|^3]. \tag{41}$$

It is also immediate that the second term in (41) may be omitted, because by (31)

$$x^m/\|x^m\|^3 = (1/\rho_m^2)\cdot exp(J_2\cdot m_m\sigma)\cdot exp(J_2\cdot\theta_m^o)\cdot e^1, \quad \int_0^{2\pi} sin(m_k\sigma)\cdot exp(J_2\cdot m_m\sigma)\, d\sigma \equiv 0 \tag{42}$$

because of the hypothesis that $m_k \neq m_m$. Thus we have simplified the calculation to that of evaluation of

$$f^{km} = \int_0^{2\pi} sin(m_k\sigma)\cdot\Psi(\sigma)\cdot(x^m(\sigma) - x^k(\sigma))\, d\sigma, \quad x^k(\sigma) = \rho_k\cdot exp(J_2\cdot[m_k\sigma + \theta_k^o])\cdot e^1, \tag{43}$$

$$\Psi(\sigma) := 1/\|x^m(\sigma) - x^k(\sigma)\|^3. \tag{44}$$

Now by the chief property of the scalar product $(x,y)$, namely that if $M$ is an arbitrary matrix $(Mx,y) \equiv (x,M'y)$, and the fact that $J_2' = -J_2$, it is easy to calculate that

$$\|x^m(\sigma) - x^k(\sigma)\|^2 \equiv \|x^m\|^2 + \|x^k\|^2 - 2\cdot\rho_m\cdot\rho_k(e^1, exp(J_2\xi)e^1), \tag{45a}$$

$$\xi = (m_k - m_m)\cdot\sigma + \theta_k^o - \theta_m^o, \tag{45b}$$

$$\Psi(\sigma) = \psi(\sigma)/[(\rho_k)^2 + (\rho_m)^2]^{3/2}, \tag{45c}$$

$$\psi(\sigma) = 1/[1 - (\varepsilon_{km})^2 cos(\xi)]^{3/2}, \quad (\varepsilon_{km})^2 := 2\cdot\rho_k\cdot\rho_m/[(\rho_k)^2 + (\rho_m)^2], \tag{45d}$$

$$\psi(\sigma) = 1 + (3/2)\cdot(\varepsilon_{km})^2 cos(\xi) + (15/8)\cdot(\varepsilon_{km})^4 cos^2(\xi) + \cdots, \tag{45e}$$

where the binomial series in (45e) always converges because $(\varepsilon_{km})^2 < 1$ whenever $\rho_k$ and $\rho_m$ are distinct (which is a consequence of the fact that then $0 < (\rho_k - \rho_m)^2$ and of obvious manipulations of the expansion of the latter).

Now, remembering that $\psi$ depends upon both $k$ and $m$, define

$$A_{kmj} := (1/[2\pi]) \cdot \int_0^{2\pi} \sin(m_k \sigma) \cdot \cos(m_j \sigma) \cdot \psi(\sigma) \, d\sigma, \tag{46a}$$

$$B_{kmj} := (1/[2\pi]) \cdot \int_0^{2\pi} \sin(m_k \sigma) \cdot \sin(m_j \sigma) \cdot \psi(\sigma) \, d\sigma, \tag{46b}$$

$$\eta_{kmj} := \{ (A_{km})^2 + (B_{km})^2 \}^{1/2}, \tag{46c}$$

$$\Phi_{kmj} := -\mathrm{Arctan}\{ B_{kmj}/A_{kmj} \}, \tag{46d}$$

and note that (43)-(44) may be simplified by use of the novel identity

$$(1/[2\pi]) \cdot \int_0^{2\pi} \sin(m_k \sigma) \cdot \psi(\sigma) \cdot \exp(J_2[m_j \sigma + \theta_j^o]) \, d\sigma \equiv \eta_{kmj} \exp(J_2[\Phi_{mkj} + \theta_j^o]). \tag{47}$$

It is the radical simplification provided by the apparently hitherto unnoticed identity (47) which appears to be the chief innovation in the present work. For we may now write that

$$f^{km} = d^{km}/[(\rho_k)^2 + (\rho_m)^2]^{3/2}, \tag{48a}$$

$$d^{km} := \left[ \rho_m \eta_{kmm} \cdot \exp(J_2[\Phi_{kmm} + \theta_m^o]) - \rho_k \eta_{kmk} \cdot \exp(J_2[\Phi_{kmk} + \theta_k^o]) \right] \cdot e^1 \equiv$$

$$\equiv \rho_m \cdot \eta_{kmm} \cdot \begin{pmatrix} \cos(\Phi_{kmm} + \theta_m^o) \\ -\sin(\Phi_{kmm} + \theta_m^o) \end{pmatrix} - \rho_k \cdot \eta_{kmk} \cdot \begin{pmatrix} \cos(\Phi_{kmk} + \theta_k^o) \\ -\sin(\Phi_{kmk} + \theta_k^o) \end{pmatrix}. \tag{48b}$$

From mere inspection of (48) it is now evident that the *necessary and sufficient conditions* for $f^{km}$ to vanish are that there be an **orbital resonance** defined by

$$\rho_m = \beta \cdot \rho_k, \quad \beta = \eta_{kmk}/\eta_{kmm}, \tag{49a}$$

and a corresponding enabling phase-shift

$$\theta_m^o = \theta_k^o + \phi_{mk} + M \cdot \pi, \quad \phi_{mk} = \Phi_{kmk} - \Phi_{kmm}, \quad (M = 1, 2, 3, \cdots). \tag{49b}$$

This completes the easy part of the present derivation.

Now begins the hard work of evaluation of $\beta$ and $\phi$. Part of this is easy, because by inspection we need only to evaluate the lowest-order terms in

$$A_{kmk} = 0 + \cdots, \quad B_{kmk} = (1/2) + \cdots, \tag{50}$$

because in this case it is adequate to use $\psi = 1 + \cdots$ because of the well known orthogonality properties of sines and cosines. In contrast, the lowest order terms in $A_{kmm}$ and $B_{kmm}$ vanish identically, and we must go to the second-order terms in $\psi$ in order to get meaningful results. Thus, to the second order in the series (45e)

$$(A_{kmm}, B_{kmm}) = (3/2) \cdot (\varepsilon_{km})^2 \cdot (\alpha_{kmm}, \beta_{kmm}) + \cdots, \tag{51a}$$

$$\alpha_{kmm} := (1/[2\pi]) \cdot \int_0^{2\pi} sin(m_k \sigma) \cdot cos(m_m \sigma) \cdot cos([m_k - m_m] \cdot \sigma + \theta_k^o - \theta_m^o) \, d\sigma, \tag{51b}$$

$$\beta_{kmm} := (1/[2\pi]) \cdot \int_0^{2\pi} sin(m_k \sigma) \cdot sin(m_m \sigma) \cdot cos([m_k - m_m] \cdot \sigma + \theta_k^o - \theta_m^o) \, d\sigma, \tag{51c}$$

which is where conceptualizing ends and labor begins. My advice to the reader is to replace each sine and cosine by the sum or difference of two complex exponentials, as in de Moivre's Theorem, and then multiply out the resulting 6 products as complex numbers. From this lengthy exercise in elementary complex algebra there results the [ultimately real] *identities:*

$$\alpha_{kmm} = -(1/4) \cdot sin(\theta_k^o - \theta_m^o), \quad \beta_{kmm} = (1/4) \cdot cos(\theta_k^o - \theta_m^o), \tag{51d}$$

from which we obtain the welcome simplification that $[ (\alpha_{kmm})^2 + (\beta_{kmm})^2 ]^{1/2} \equiv 1/4$. Therefore, finally, by (46c)

$$\eta_{kmm} = (3/2) \cdot (\varepsilon_{km})^2 \cdot (1/4) + \cdots, \quad \eta_{kmk} = (1/2) + \cdots, \tag{51e}$$

so that, by (49a), to lowest order in $\mu$,

$$1/\beta = \rho_k/\rho_m = \eta_{kmm}/\eta_{kmk} = (3/2) \cdot \rho_k \cdot \rho_m /[ (\rho_k)^2 + (\rho_m)^2 ]^{3/2} \equiv$$

$$\equiv (3/2) \cdot (1/\beta)/[ 1 + (1/\beta)^2 ]^{3/2}, \tag{51f}$$

which *requires* for self-consistency that, reminiscent of the **1766/1772 Titius/Bode 'Law'**,

$$1 + (1/\beta)^2 = (3/2)^{2/3}, \tag{51g}$$

i.e. that $\beta$ have the *unique* particular value claimed in (30b)! Note that to lowest order this $\beta$ is a *universal constant*, **independent** of the gravitational constant $G$ or the masses of the planets!

The proof of (30a) is analogous but simpler, noting that from (46d) and (51d)

$$\phi_{mk} = \Phi_{kmk} - \Phi_{kmm} = -(\theta_m^o - \theta_k^o), \tag{51h}$$

so that, bringing $(\theta_m^o - \theta_k^o)$ to the left-hand side of (49b) and dividing by 2 we obtain the claimed result (30a).

Equation (38) is a trivial consequence of (48b).  Also, evaluation of the Jacobian with respect to ($\rho_k$, $\theta_k^o$) instead of with respect to the components of $x^k(0)$ makes no difference to the claim that the Leray-Schauder Index of the generating solution isolated by (51g) is unity.  This completes the proof.  □

## Relationship to Roy & Ovenden's Resonances

Note carefully that the preceding mathematically rigorous result applies only to a highly idealized and limited physical model of the planetary system, specifically to the partially decoupled & hierarchical planar problem.  Its main physical significance seems to be qualitative, in that it has been proved that for *certain* solar systems one may expect a Titius-Bode Law to adhere rather precisely.  However, if one broadens the problem from consideration only of 3-body interactions to allow simultaneous 4-body and  even 5-body interactions, there is no reason to doubt that similar results could be proved.  In this way the presently observed Resonant Structure of the solar system can be "explained" as obviously of dynamical rather than cosmogonical origin.

During the 1950's Roy & Ovenden (*cf.* refs. in [5]) and in 1966 Molchanov called attention to the fact that the nine planets fit closely to a resonant pattern.  Following the latter, let the 9-vector $\vec{\omega} = (\omega_i) \in \mathbb{E}^9$ denote a set of Newcomb-type frequencies in a planar formulation, and let the matrix $\mathbf{M} = (M_{ij})$ denote the following $8 \times 9$ matrix:

$$\mathbf{M} = \begin{pmatrix} 1 & -1 & -2 & -1 & 0 & 0 & 0 & 0 & 0 \\ 0 & 1 & 0 & -3 & 0 & -1 & 0 & 0 & 0 \\ 0 & 0 & 1 & -2 & 1 & 1 & 1 & 0 & 0 \\ 0 & 0 & 0 & 1 & -6 & 0 & -2 & 0 & 0 \\ 0 & 0 & 0 & 0 & 2 & -5 & 0 & 0 & 0 \\ 0 & 0 & 0 & 0 & 1 & 0 & -7 & 0 & 0 \\ 0 & 0 & 0 & 0 & 0 & 0 & 1 & -2 & 0 \\ 0 & 0 & 0 & 0 & 0 & 0 & 1 & 0 & -3 \end{pmatrix}. \tag{52}$$

Then it will be true to a high degree of approximation that

$$\mathbf{M} \cdot \vec{\omega} = 0, \tag{53}$$

where most of the errors on the right-hand side, which should be exactly 0, are smaller than 0.003 and the largest in absolute value is $|-0.012| = 0.012$.

The Asteroid Belt is not included, but it could have been shown  --  with reference to a hypothetical *missing planet* **Astex** [similar to Ovenden's Aztex] as a 3:1 resonance $1 \cdot \omega_A - 3 \cdot \omega_J = 0$ with Jupiter.  The matrix does show the famous $2 \cdot \omega_J - 5 \cdot \omega_S = 0$ resonance betweeen Jupiter and Saturn.  If we replace the $1 \cdot \omega_J - 7 \cdot \omega_U = 0$ resonance between Jupiter and Uranus by the logical consequence of utilization of the previous resonance we get a resonance $5 \cdot \omega_S - 14 \cdot \omega_U = 0$ which can be taken to be illustrative of the presently proved (30d) if we had allowed integers larger than 5, and, moreover, agrees with (30d) in the form $1 \cdot \omega_S - 3 \cdot \omega_U = 0$ if we make the 7% error of rounding off 2.80 to 3, while the discrepancy with (30d) can be ascribed to the missing higher-order terms in the power-series in the small parameter µ. Finally the $1 \cdot \omega_U - 2 \cdot \omega_N = 0$ from the matrix does fit (30d) perfectly. In short, subject to justification of inclusion of Astex, which will be done later, it is evident that in the actual Solar System the 4 planets

( **Astex**, **Jupiter**, **Saturn**, **Uranus**, **Neptune** )

are [to first order in µ] in accord with the present theory (30d) by virtue of the *well-known* **resonances**

$$1 \cdot \omega_A - 3 \cdot \omega_J = 0, \quad 2 \cdot \omega_J - 5 \cdot \omega_S = 0, \quad 1 \cdot \omega_S - 3 \cdot \omega_U = 0, \quad 1 \cdot \omega_U - 2 \cdot \omega_N = 0. \tag{54}$$

It seems intuitively evident that if one included complications omitted from the present admittedly idealized, simplified theory (such as elliptical generating solutions, non-zero inclinations, and 4-body and 5-body interactions) then a theorem similar to that above could be proved which would account for the resonant structure of the entire planetary system. However, there is an analytically more convenient approach to the many-body problem, which obtains similar results, and to which we now turn.

## Relationship to Ovenden's Least-Interaction Action and Bass's Least Mean Absolute Potential Energy

In 1958 at the International Congress of Mathematicians in Edinburgh I distributed informally a Martin-Marietta internal report which was again in 1960 presented officially (and printed in the *Proceedings* Abstracts) at the *IAF* in Stockholm [35] which *inter alia* contained the following result, that I called the *Principle of Least Mean Absolute Potential Energy*. "The 'least unstable' almost periodic orbits of any homogeneous, definite, conservative dynamical system, for each compatible value of the energy constant, are precisely those closed curves in configuration space along which the average of the [absolute value of the] system's potential energy has a minimum value (relative to all nearby admissible closed curves). ... This principle is a generalization of the Lagrange-Dirichlet *Principle of Least Potential Energy*. ... The minimization should be done as follows: (a) fix angular momentum; (b) fix Mean Value $<KE> = \mathcal{E}_0 > 0$ of Kinetic Energy along trial path; (c) vary hypothetical path in $6 \cdot n$-dimensional phase space until the Mean Value $<|V|>$ of the absolute potential energy has a stationary (critical) value $\mathcal{E}_{MIN} > 0$; (c) adjust motion until the Virial Theorem is satisfied, i.e. until $\mathcal{E}_{MIN} = 2 \cdot \mathcal{E}_0$. Then the corresponding motion actually satisfies Newton's equations for the gravitational *n*-body problem." (An *autosynartetic* solution [41] is one in which the condition of periodicity $x(T) = x^0$ is replaced by $x(T) = U \cdot x^0$ where *U* is an orthogonal matrix, and similarly for $v(T) = dx/dt$; the point is that in the boundary condition which is key here to use of the variational calculus, the usual identity between the scalar product $x(T) \cdot v(T)$ and $x^0 \cdot v^0$ is preserved because $Ux(T) \cdot [Uv(T)] \equiv x(T) \cdot [(U'U)v(T)]$ and $U'U$ is the identity matrix.)

In 1970 J.G. Hills [36] published computer simulations of hypothetical planetary configurations exhibiting initial vigorous interaction after which (to quote Roy [5]) "... all the systems tended toward quasistationary states, each state showing a tendency for the periods of adjacent orbits to be near-commensurable in small integer fractions." Then Hills concluded that "this suggests to me that Bode's Law may have resulted from a process of **dynamical relaxation**."

In 1972-75 (see [37] plus multiple further references in [5], dedicated by Roy to the memory of Ovenden) the late dynamical astronomer Michael W. Ovenden pursued and confirmed Hill's idea with additional computer simulations (although Roy points out that the numerical accuracy & techniques employed leave much to be desired) which led him to conjecture his *Principle of Least Interaction Action*: "A planetary or satellite system of N point masses moving solely under their mutual gravitational attractions spends most of its time close to a configuration for which the time mean of the action associated withe the mutual interactions of the planets or satellites is a minimum."

After verifying his Principle for the known satellite systems, Ovenden noted that the present Solar System is not quite at a minimum, but would be if one postulated a planet of some 90 Earth masses in the place of the asteroid belt [37]. He called this hypothetical missing planet Aztex. Astronomer Tom Van Flandern [38] has marshalled an impressive array of evidence for the former existence of such a planet (which he playfully calls Planet K after the modern myth of Krypton), though he would postulate a smaller planet than Aztex.

I recently learned that in one of Ovenden's later papers he called attention to the obvious similarity between his Principle and my own, while graciously acknowledging my priority.

During 1995 I have attempted, in work to be reported in more detail elsewhere, to refine Ovenden's calculations by doing them analytically in order to avoid the very just criticism of Roy as regards their numerical imperfection. By using the power-series in μ developed above, and carrying enough terms to get a definitive answer, it is possible to arrive at analytical expressions for the Mean Interaction Action of *n* planets because the assumed resonance [as in the Roy-Ovenden/Molchanov analysis (52)-(53) of the present solar system] eliminates the "small divisor" problem and makes it possible to evaluate the integrals in closed form via Elliptic Functions. My results are that the present solar system is at a local Interaction-Action minimum if one postulates a planet in the asteroid belt with only 2.045 times the Earth's mass. Accordingly, to distinguish my twice-Earth sized missing planet from Ovenden's Saturn-sized Aztex, I am calling my 'missing planet' **Astex**, which I presume to have orbited at a distance of 2.65 AU from the Sun.

Space precludes a presentation here of the detailed derivation of the mass of Astex.

However, the reader may find of interest a summary of research which I did on an arbitrary number $n$ of bodies, and presented in a seminar at the University of Alabama in Hunstville (UAH) during November, 1992, after numerical assistance by UAH Prof. C.D. Johnson and his students, particularly Stuart Addington of Teledyne Brown, Inc.

Now that the importance of the distal multiplier $\beta$ for the distance beyond an inner planet for an outer planet in terms of the ratio of orbital radii has been derived in full rigor for the coplanar, concentric 3-body problem, it seems permissible to use a cruder physical model to consider the case of 4 or more bodies. For simplicity, keep the first body anchored at the origin and let the bodies have mean orbital radii $\rho_i$ ($i = 1, 2, \cdots n$), where $\rho_1 \equiv 0$ by definition, and

$$0 = \rho_1 < \rho_2 < \cdots < \rho_i < \rho_{i+1} < \cdots < \rho_n, \tag{55}$$

and where each body (initially assumed decoupled from mutual interactions) is started with Keplerian circular velocity $v_i = \sqrt{G \cdot M}/\sqrt{\rho_i}$ where $M = m_1$ denotes the large central mass and where the smaller masses are denoted by $m_i = \mu_i \cdot M$ in terms of ratios $\mu_i \ll 1$ ($i = 2, 3, \cdots, n$). Let $\Gamma_0 > 0$ denote initial total angular momentum, and $E_0 = |\mathcal{E}_0| = -\mathcal{E}_0 > 0$ denote the absolute value of the initial total energy. Also, define $\gamma_0 \equiv \Gamma_0/\sqrt{G} \cdot M^{3/2}$ and $\varepsilon_0 \equiv 2 \cdot E_0/\sqrt{G} \cdot M^2$, and it is easy to verify that (neglecting terms quadratic or higher in the $\mu_i$) Conservation of Angular Momentum and Conservation of Energy (combined with the Virial Theorem) are given by $\gamma = \gamma_0$ and $\varepsilon = \varepsilon_0$ where

$$\gamma = \sum_{i=2}^{n} \mu_i \sqrt{\rho_i}, \qquad \varepsilon = \sum_{i=2}^{n} \mu_i/\rho_i, \tag{56}$$

THEOREM. *If the planetary orbits satisfy a Titius-Bode Law of the form*

$$\rho_i = \beta^{i-2} \cdot \rho_2, \quad (i = 2, 3, 4, \cdots, n), \tag{57}$$

*then the distal ratio $\beta = z^2$ is the square of the unique positive root $z > 0$ of a polynomial of degree $4 \cdot n - 8$ in $z$ whose coefficients are functions only of the constants $\gamma_0$ and $\varepsilon_0$ and the mass-ratios $\mu_i$ and which can be derived by elimination of $\rho_2$ between the expressions for $\gamma$ and $\varepsilon$ in (56).*

PROOF. Insert (57) into (56) and compute $\gamma^2 \cdot \varepsilon$. I shall publish a general algorithm defining all of the coefficients explicitly elsewhere; however, it can be recovered easily by the interested reader after following the next example in the case $n = 3$, wherein the algebra is less difficult. □

REMARK. To apply this result to the present solar system, simply replace the $\rho_i$ in (56) by the actual mean distances $R_i$ from observational astronomy and use the actual mass-ratios $\mu_i$ from astrophysics in order to evaluate $\gamma_0$ and $\varepsilon_0$ numerically. After finding $\beta$, the eliminated $\rho_2$ can be recovered by solving either $\gamma = \gamma_0$ or $\varepsilon = \varepsilon_0$, following which the remaining $\rho_i$ are given by (57).

In the case $n = 3$, the polynomial $F(z)$ whose roots are sought has the form

$$F(z) = z^4 + \alpha_3 \cdot z^3 + \alpha_2 \cdot z^2 + \alpha_1 \cdot z^3 + \alpha_0, \tag{58}$$

which, as in the proof of the preceding theorem, can be obtained simply by comparing coefficients in the multiplied-out version of

$$(\mu_1 \cdot z^2 + \mu_2) \cdot (\mu_1 + \mu_2 \cdot z)^2 - \alpha \cdot z^2 = 0, \quad \alpha = \varepsilon_0 \cdot (\gamma_0)^2. \tag{59}$$

The result is

$$\alpha_3 = 2 \cdot (\mu_1/\mu_2), \quad \alpha_2 = (\mu_1/\mu_2)^2 + (\mu_2/\mu_1) - \alpha/[\mu_1(\mu_2)^2], \quad \alpha_1 = 2, \quad \alpha_0 = \mu_1/\mu_2. \tag{60}$$

Using data for Jupiter and Saturn, the reader may verify that $F = 0$ has only one positive root, whosee square yields $\beta = 1.833$. With more effort, analogous results can be obtained for arbitrary $n$, which I sent to Prof. C.D. Johnson of UAH in 1992; I am exceedingly grateful to him and to Stuart Addington for evaluating these formulas for each $n$ between $n = 2$ and $n = 11$ in the present Solar System (augmented by the hypothetical Astex). In 1992 we were wavering between Ovenden's massive Aztex and other evidence (a least-mean square fit of the Solar System not described here) suggesting that $\mu_6$ is about three times the Earth's value of $\mu_4 = 3.003 \times 10^{-6}$ (as I had not yet found the refined estimate of 2.045 times the Earth's mass described above), and so we used 3.3477 Earth masses for Astex. Our results were:

| $m_i$ | $R_i$ | $\rho_i$ | **BODY** | $n$ | $\beta$ | $\beta_i$ |
|---|---|---|---|---|---|---|
| 1,000,000 | | | Sun | 1 | | |
| 0.16601 | 0.3871 | 0.2818 | +Mercury | 2 | | |
| 2.447841 | 0.7233 | 0.50608 | +Venus | 3 | 1.86 | 1.868 |
| 3.003469 | 1.000 | 0.9088 | +Earth | 4 | 1.46 | 1.382 |
| 0.322714 | 1.5237 | 1.632 | +Mars | 5 | 1.47 | 1.5237 |
| 10.054798 | 2.7 | 2.93 | +Astex | 6 | 1.58 | 1.772 |
| 954.60258 | 5.2030 | 5.26 | +Jupiter | 7 | 1.68 | 1.927 |
| 285.80765 | 9.5281 | 9.45 | +Saturn | 8 | 1.766 | 1.831 |
| 43.549846 | 19.1829 | 16.97 | +Uranus | 9 | 1.800 | 2.013 |
| 51.67161 | 30.0796 | 30.48 | +Neptune | 10 | 1.795 | 1.568 |
| 0.007541 | 49.0250 | 54.70 | +Pluto | 11 | 1.795 | 1.6298 |

**Table 1**

Note that, by the stage at which the full present Solar System had been included, the value of β had "converged" to β = 1.795, in amazing agreement with the more rigorously derived $β_0$ of (30b) above!

**Plausibility From Computer-Simulation Experiments**

The skeptical reader can readily verify the plausibility of the present main results by performing the following computer-simulation experiments.

First, consider a solar system with only the Sun, Jupiter and Saturn. Start the two planets on opposite sides of the Sun with Keplerian circular velocities and integrate numerically for 100 years. In this time, Jupiter will make 8.43 complete revolutions while Saturn will make just 3.4 revolutions around the Sun. Using a Runge-Kutta Fourth-Order integrator in compiled QuickBasic on a 133 MHz PC and a time-step of less than one day, the results displayed on a VGA monitor show *perfect circles* with not even a single pixel's discrepancy after multiple revolutions around the origin.

Next, keep everything the same, but move Jupiter say 30% closer to Saturn. If there is nothing special about the Bode's Law relationship of the previous simulation, then Jupiter and Saturn should again describe perfect circles for 100 years. Instead, after just 30 years the orbit of Saturn *visibly* fails to close upon itself, and after Saturn has made 3 complete revolutions the trail of previous points shows that each orbit has been an erratic spiral with each body *visibly* destabilizing the formerly orderly motion of the other throughout the entire century.

One way to understand the destabilization is to note with Chambers *et al* [32] that according to Gladwin's criterion the distance between adjacent concentric orbits must be more than $2\cdot\sqrt{3}$ measured in units of mutual Hill radii in order to avoid an instability of the near-collision type; and by numerical examples they raise Gladwin's limit to about 10. However, in the present Solar System only the distance between Jupiter and Saturn is even near to the border set by the latter criterion; but in moving Jupiter outward, we crossed that border. In fact, the distance $D$ in mutual Hill radii between the present planetary orbits can be computed to be: $D$(Mercury,Venus) = 63.4; $D$(Venus,Earth) = 26.3; $D$(Earth,Mars) = 40.1; $D$(Mars,Jupiter) = 16.0; $D$(Jupiter,Saturn) = 7.895; $D$(Saturn,Uranus) = 14.0; $D$(Uranus,Neptune) = 13.95; and $D$(Jupiter,Saturn) is indeed the smallest.

An even more striking visual demonstration can be provided for the benefit of doubters that there is anything special about the Solar System's present configuration. First, using the same procedure just mentioned, I integrated the entire Solar System (omitting Mercury and Uranus and Neptune) for 200 years to produce *perfect* concentric circles. But then I placed Venus at 0.5 AU, kept Saturn at its present distance, and placed the other 4 planets *equidistantly* between these inner and outer bodies, producing a "harmonically-space" configuration. When I integrated this harmonic solar system for 200 years, visible manifestations of resonant destabilization were very soon evident in all 6 orbits!

In my opinion, these simple computer experiments provide a strong "plausibility proof" that the Solar System's present resonant configuration is indeed the result of J.G. Hill's **"dynamical relaxation"** in which a positive-definite functional of the Bass-Ovenden type is minimized.

The idealized Newtonian gravitational model is non-dissipative and theoretically reversible. However any actual planetary system is subject to non-Hamiltonian forces which, no matter how small they may seem, MUST be taken into account when the **extreme** *sensitivity to initial conditions* ("dynamical chaos") of Hamiltonian systems is considered. To quote myself by quoting Roy's paraphrase of my 1975 paper [21]: "... the work of J.G. Hills and M.W. Ovenden demonstrated that, after a short period of wild behavior, a planetary system could settle down into a distribution of orbits very similar to a commensurable Bode-type configuration. Such a configuration would, under the action of other forces such as tidal friction, nudge the system into a neighboring truly stable configuration, which on inspection might be thought to have been the system's state for a very long time. Indeed, numerical integrations backward in time could take the system, still well behaved, through the episode of wild behavior as if it had never been."

Accordingly it was sheer **incompetent folly** for the critics of such theories of recent catastrophic planetary interactions as those of the late Immanuel Velikovsky and his successors to assert dogmatically [41] that dynamical astronomy "proves" that such scenarios are "physically impossible."

## Conclusions

The principal result above does not depend upon the resonant commensurability $m_i:m_{i+1}$ involving only *small* relatively prime integers $m_k$ ( $k = i, i + 1$ ); indeed, these integers can be *arbitrarily large*, which places a totally different perspective on the main result above. Thus, an alternative formulation of the main result, assuming the reader's familiarity with the advanced probabilistic ('measure theoretic') notions of Appendix 1 below, goes as follows:

<u>DEFINITION</u>. *A result applies to an initial configuration* **with probability unity** *if it applies rigorously after the adjustment of the ICs by an* **arbitrarily small** *amount.*

<u>REMARK</u>. Thus two concentric circles about the center of gravity of the Sun and two very much smaller planets having as radii the mean distances from the Sun to Jupiter and Uranus, respectively, could [as the masses of the planets are made "sufficiently" small] have a property with *probability unity* provided that the property holds when the radii are adjusted by amounts **smaller than the radius of a proton**!)

<u>THEOREM</u>. *Consider the problem of isoperiodic continuation to some* $\mu > 0$ *of the initially decoupled* ( $\mu = 0$ ) *Keplerian circular orbits of the planar* 3*-body problem, wherein the concentric circles defining the orbits at* $\mu = 0$ *of the two small planets have completely ARBITRARY radii. Then* *with probability unity the necessary and sufficient condition for the existence of a periodic orbit for either planet for any sufficiently small value of* μ > 0 *is that the initial radii of the two orbits at* μ = 0 *should have the precise ratio*

$$\beta_o \;=\; 1/\sqrt{(3/2)^{2/3} - 1} \;=\; 1.80\ldots\;, \tag{61}$$

*where the new* **natural constant** $\beta_o$ *is INDEPENDENT of both the Newtonian gravitational constant and the initial masses of the Sun and small planets.*

It is commonly believed that rectangular coordinates are not well-adapted to the Newtonian *n*-body problem, which is why myriads of other coordinate systems have been introduced. Nevertheless, it seems to me that it is precisely because I remained in the Cartesian coordinates that the powerful results about the infinitesimal generators of rotation groups which were the key elements in enablement of the above analytical derivation became available.

During 1953-54 I had the privilege of attending the late Carl Ludwig Siegel's *Vorlesungen über Himmelsmechanik* sponsored at JHU by my late teacher Aurel Wintner prior to their publication. Seeing Siegel at work reminded me of a Bulldozer; he never went 'around' a problem, he just ploughed straight ahead and eventually ploughed right on through it. The above utterly straightforward approach to a long-standing mystery of celestial mechanics, if found to be correct, may be an echo of the Siegel approach.

## Acknowledgments

For primary encouragement, which led to my first foray [39] into this subject, and for the help in generation of Table 1 by extensive calculations, I am deeply indebted to Dr. C.D. Johnson and to his former student Stuart Addington.

As already mentioned, this work could never have been done without the encouragement and support of David Talbott, publisher of *Pensée*, and Charles Ginenthal, publisher of *The Velikovskian*. Also I am indebted to Dr. Miklos Farkas [15], to Dr. John E. Chambers [33], and to Dr. Philip Morrison [32], for their interest and for valuable literature references.

## Appendix 1

A positive real number number $r \in [0,1]$ is *rational* if it is a fraction of whole numbers, i.e. if $r = p/q$ where $p$ and $q$ are relatively prime positive integers (whole numbers). In decimal notation, a rational number is either a terminating expression, such as $1/10 = 0.1$ or an endlessly repeating decimal, such as $1/3 = 0.33333\cdots$. If the decimal fraction expression does not repeat, then the number is irrational. The main distinction between rational numbers and irrationals is that the former are 'countable' or 'denumerable' while the latter are 'uncountably infinite' and have, in the sense of Cantor, the 'cardinality of the continuum.' Cantor's Matrix is an infinite array which lists every possible fraction by the simple expedient of placing the fraction i/j in the *j*th column of the *i*th row; then, starting at the upper left-hand corner, and preceding along ever-longer counter-diagonals (lines of 45 degrees slope rising from left to right) one can 'count' or 'denumerate' every rational number as an element of a *sequence* { $r_m$ | m = 1, 2, 3, ⋯ }. The 'length' of the portion of the line-segment [0,1] taken up by all the rational numbers is zero, in the sense that we can cover each entry in the sequence { $r_m$ } with a sequence of non-overlapping line-segments of lengths each not exceeding $\eta/2^m$, and then add up the lengths of all these covering-segments to get the geometric series

$$\eta/2 + \eta/4 + \eta/8 + \cdots = (\eta/2)\cdot\left(1 + \sum_{m=1}^{+\infty} 2^{-m}\right) \equiv (\eta/2)\cdot[1/(1-2^{-1})] \equiv \eta,$$

where η was arbitrarily small to begin with. Therefore in the sense of probability theory the 'length' of the segment [0,1] taken up by the rationals is zero, and so the 'length' of the complementary irrationals must be 1.

Some non-mathematicians are tempted to say that since the probability of a resonance occurring is 'zero', the resonant orbits can be ignored. This is fallacious for two reapsons. In the first place, in probability theory, events can occur which have probability zero. In the second place, the number of planets is finite, and so *counting* is not 'non-physical'! Therefore, assuming that one planet orbits the barycenter *p* times while another orbits it *q* times is not 'non-physical'; this amounts to the assumption that there is no relative precession in the line of apsides, which is characteristic of resonant orbits. As Einstein showed (resolving a centuries old anomaly regarding the precession of the perihelion of Mercury by 43 seconds of arc per century), the generalization of Newtonian Gravitational Mechanics to General-Relativistic Gravitational Mechanics will introduce precession into even two-body problems; however, in our solar system this effect is negligible except for Mercury.